\documentclass[%
 aip,
 rsi,
amsmath,amssymb,
reprint,
floatfix
]{revtex4-1}
\usepackage{graphicx}
\usepackage{multirow}
\usepackage{color} 

\draft 

\newcommand{\PNNL}{%
   \affiliation{%
       Physical Sciences Division, Pacific Northwest National Laboratory, Richland, WA 99354, USA
       }}

\begin{document}

\title{Quantum simulations employing connected moments expansions}

\author{Karol Kowalski} 
\email{karol.kowalski@pnnl.gov} \PNNL
\author{Bo Peng} 
\email{peng398@pnnl.gov} \PNNL
\date{April 2020}

\date{\today}

\begin{abstract}
Further advancement 
of quantum computing (QC) is contingent on enabling many-body models 
that avoid deep circuits and excessive use of  CNOT gates. To this end, we develop a QC approach employing 
finite-order connected moment expansions (CMX) and affordable procedures for initial state preparation. We demonstrate the performance of our approach employing several quantum variants of CMX through the classical emulations on the H$_2$ molecule potential energy surface and the Anderson model with a broad range of correlation strength.  The results show that our approach is robust and flexible. Good agreements with exact solutions can be maintained even at the dissociation and strong correlation limits.


\end{abstract}

\maketitle

\section{Introduction} 
The quantum computing (QC) techniques attract 
much attention in many areas of mathematics, physics, and chemistry by providing a means to address insurmountable computational barriers  for 
simulating quantum systems on classical computers. 
One of the best illustrations of this fact  is associated with solving Schr\"odinger equations for many-electron systems in quantum chemistry, where the attempt of including all configurations  spanning Hilbert space (or at least configurations from the subspace relevant to a problem of interest)
quickly evolves into numerical problems characterized by  exponentially growing  numerical cost. 
Unfortunately, for a large class of the problems commonly referred to as the strongly correlated systems, the inclusion of all configurations  is necessary 
to obtain  a desired level of accuracy.
To bypass these problems, several classes of many-body methods targeting the so-called full configuration interaction (FCI)  accuracy have been developed including high-rank coupled-cluster (CC),\cite{cizek66_4256,paldus07,bartlett_rmp}
selected CI,\cite{tubman2016deterministic,liu2016ici}
density matrix renormalization group (DMRG), 
\cite{white1992density,schollwock2005density,dmrg2,chan2011density}
stochastic and semi-stochastic CI,\cite{booth2009fermion,booth2013towards,petruzielo2012semistochastic}
hybrid stochastic CI and CC,\cite{deustua2017converging,deustua2018communication,yuwono2020accelerating}
and  virtual orbitals many-body expansions \cite{eriksen2017virtual}  methodologies. Recently,  accuracies of these formalisms have been evaluated on the example of  benzene system, \cite{eriksen2020ground} where considerable differences in the calculated energies with high-accuracy formalisms were reported. 

Although QC has not reached its maturity yet, intensive effort has been directed towards developing algorithms for quantum simulations  that can take advantage of early noisy quantum registers. 
Among  several algorithms that are being developed for this purpose one should mention hybrid quantum/classical Variational Quantum Eigensolver (VQE) approach,
\cite{peruzzo2014variational,mcclean2016theory,romero2018strategies,PhysRevA.95.020501,Kandala2017,kandala2018extending,PhysRevX.8.011021,huggins2020non}
that  has been intensively tested from the point of view of its 
utililzation on noisy intermediate-scale quantum devices (NISQ).
Other quantum algorithms such as 
quantum phase estimation (QPE),
\cite{luis1996optimum, cleve1998quantum,berry2007efficient,childs2010relationship,seeley2012bravyi,
wecker2015progress,haner2016high,poulin2017fast} 
and more recently 
quantum algorithms for imaginary time evolution (ITE),\cite{mcardle2019variational,motta2020determining} 
quantum filter diagonalization,\cite{parrish2019quantum} quantum inverse iteration algorithms,\cite{kyriienko2020quantum} and quantum power methods \cite{seki2020quantum} 
seem to be more hardware demanding than the VQE algorithm. 
%
Some of the above mentioned algorithms may
draw heavily on the utilization of Trotter formulas for evaluating the exponential forms of the operators corresponding  to unitary coupled cluster ansatze, unitary time evolution of the system, and the solution of the imaginary-time Schr\"odinger equation, respectively. 
For example, for the unitary time evolution defined by qubit-encoded Hamiltonian $H$
(various types of encoding algorithms for many-body Hamiltonians  - including Jordan-Wigner \cite{jordan1993paulische} and Bravyi-Kitaev \cite{bravyi2002fermionic}  methods - are discussed in Refs.\cite{seeley2012bravyi,mcardle2020quantum}),
where $H$ is sum of multiple (generally non-commuting) terms 
$H_j$,  $H=\sum_{j=1}^{M} H_j$ (here $H_j=h_j P_j$ with $h_j$ being complex numbers and $P_j$ representing tensor products of Pauli matrices and/or identity matrices), it is hard to encode 
$e^{-iHt}$ directly into quantum gates. 
Instead in quantum simulations  one resorts to the Trotter formula \cite{trotter1959product,suzuki1990fractal} with $K$ Trotter steps:
\begin{equation}
e^{-iHt}\simeq [U(\Delta_t)]^K
\label{trsu}
\end{equation}
where $\Delta_t=t/K$ and 
for one Trotter step $U(\Delta_t)$ is further approximated as $\prod_{j=1}^{M} e^{-i h_j P_j \Delta_t}$.
Note that even in its simplest case ($K=1$) Eq. (\ref{trsu}) results in arbitrary connectivity between qubits. With number of terms in Hamiltonian being proportional to $\mathcal{O}(N^4)$ (where $N$ stands for the number of one-particle basis functions employed) 
the optimization of the corresponding circuits plays a crucial role in the quantum simulations (especially when employing the techniques related to the fermionic swap network\cite{low_depth_Chan}).
Nevertheless,  even for formulations using  reduced  number of terms (e.g. proportional to $\mathcal{O}(N^2)$) in many-body Hamiltonian, the resulting gate depth still depends polynomially on $N$ and the total number of CNOT gates is still proportional to $\mathcal{O}(N)$ (letting alone the multiple CNOT gates for multi-qubit operators needed for example in QPE). 

In this letter we propose an alternative approach for  quantum simulations of many-body systems, where the number of instances for using the Trotter formula is  significantly reduced.
For this purpose we  utilize connected moments expansion introduced by Horn and Weinstein\cite{horn1984t} and further advanced by Cioslowski \cite{cioslowski1987connected} and others \cite{knowles1987validity,stubbins1988methods,perez1988t,yoshida1988connected,ullah1995removal,mancini1994analytic,mancini1997numerical,zhuravlev2016cumulant}, 
where the energy of a quantum system can be calculated using trial wave function $|\Phi\rangle$ (non-orthogonal to the exact ground state $|\Psi\rangle$) and the expectation values of the Hamiltonian powers $\langle\Phi|H^n|\Phi\rangle$. Here, $\langle\Phi|H^n|\Phi\rangle$'s are calculated using simple variants of Hadamard test, where
 CNOT gates are used to  control the products  of the unitaries $P_j$. Additional CNOT gates are optionally required for strongly correlated systems  where correlation effects need to be captured in the process of trial state preparation, which in a realistic setting requires only the inclusion of a relatively  small number of parameters in the VQE formulations. 
 We  also evaluate the effect of noise on the accuracy of quantum CMX simulations. 
 \\
%
%
\section{Connected moments expansion} 
The connected moments expansion is derived from  Horn-Weinstein (HW) theorem  \cite{horn1984t} that relates the function $E(\tau)$  to a series expansion in $\tau$ with coefficients being represented by connected moments $I_k$,
\begin{equation}
    E(\tau)=\frac{\langle\Phi|He^{-\tau H}|\Phi\rangle}
    {\langle\Phi|e^{-\tau H}|\Phi\rangle}=\sum_{k=0}^{\infty} \frac{(-\tau)^k}{k!} I_{k+1}
    \;.
    \label{hw1}
\end{equation}
Here the connected moments $I_k$ are defined through a recursive formula (see Ref.\citenum{horn1984t} for details)
\begin{equation}
    I_k=\langle\Phi|H^k|\Phi\rangle - \sum_{i=0}^{k-2}
    {k-1 \choose i} I_{i+1} \langle\Phi|H^{k-i-1}|\Phi\rangle.
    \label{hw3}
\end{equation}
It can be shown that $E(\tau)$ is monotonically decreasing and 
$\lim_{\tau\rightarrow\infty} E(\tau)$ corresponds to the exact ground-state energy $E_0$, which is a consequence of the fact that $e^{-\tau H}$
contracts any trial wave function $|\Phi\rangle$ towards the true ground state $|\Psi\rangle$ (assuming $\langle\Phi|\Psi\rangle\ne 0$). In practical applications based on the truncated moments expansion, Pad\'e approximants need to be used to reproduce the proper behavior in the
$\tau=\infty$ limit. The behavior of this expansion  and some techniques  for  maintaining size-extensivity were discussed in Ref.\citenum{stubbins1988methods}
(see also Ref.\citenum{piecuch2003exactness}). 

By applying re-summation techniques to Eq. (\ref{hw1}) Cioslowski  \cite{cioslowski1987connected} derived an analytical  form for exact energy in the $\tau=\infty$ limit
\begin{eqnarray}
    E_0=I_1-\frac{S_{2,1}^2}{S_{3,1}}(1+\frac{S_{2,2}^2}{S_{2,1}^2S_{3,2}}\ldots (1+\frac{S_{2,m}^2}{S_{2,m-1}^2S_{3,m}})\ldots)), \nonumber\\
 \label{hw4}
\end{eqnarray}
where $S_{k,1}=I_k~~(k=2,3,\ldots)$ and $S_{k,i+1}=S_{k,1}S_{k+2,i}-S_{k+1,i}^2$.
The truncation of the CMX series (\ref{hw4}) after the first K terms leads to the CMX(K) approximations. For example, CMX(2) and CMX(3) energies are defined as follows,
\begin{eqnarray}
    E_0^{\rm CMX(2)}&=& I_1 - \frac{I_2^2}{I_3} \;,\;\;\label{cmx2} \\
    E_0^{\rm CMX(3)}&=& I_1 - \frac{I_2^2}{I_3}
    -\frac{1}{I_3} \frac{(I_2I_4-I_3^2)^2}{I_5I_3-I_4^2}\;.\label{cmx3}
\end{eqnarray}
The CMX formalism provides a  trade-off between the rank of the connected moments included in the approximation and the quality of the trial wave function. 
Specifically, it allows one to use the multi-configurational \cite{cioslowski1987connected} or even 
correlated representations (in the form of truncated CI or CC wave functions \cite{noga2002use}) of the trail wave functions for quasi-degenerate sought-for-states. 

The analytical properties of  CMX expansions and their relations to Lanczos methods have been discussed in the literature.
\cite{knowles1987validity,prie1994relation,mancini1994analytic,ullah1995removal,mancini1995avoidance,fessatidis2006generalized,fessatidis2010analytic} 
Several techniques have been introduced to counteract possible problems associated with singular behavior of the CMX expansions (as pointed out by Mancini there exists an infinite number of valid CMX expansions \cite{mancini1994analytic}) including Knowles's generalized Pad\'e approximation,\cite{knowles1987validity} alternate moments expansion (AMX),\cite{mancini1994analytic} and generalized moments expansion.\cite{fessatidis2006generalized}
Additionally, as suggested by Noga {\it et al.}  in Ref. \citenum{noga2002use}
a proper definition of the trial wave function may significantly alleviate possible  issues associated with singular behavior. 
An interesting yet less known CMX formulation has  been proposed by
Peeters and Devreese \cite{peeters1984upper} (further analyzed by
Soldatov
\cite{soldatov1995generalized}) where the upper bound of the ground-state energy in $n$-th order approximation is calculated as a root of the polynomial $P_n(x)$
\begin{equation}
    P_n(x) = \sum_{i=0}^n a_i x^{n-i} \;, 
    \label{hw6}
\end{equation}
where
$a_0=1$
and
coefficients $a_i$'s for $1\le i \le n$
(forming vector ${\bf a}$) are obtained by solving linear equations
\begin{equation}
    {\bf M}{\bf a}=-{\bf b}
\end{equation}
with matrix elements  ${\bf M}_{ij}=\langle\Phi|H^{2n-(i+j)}|\Phi\rangle$ and vector component 
$b_i=\langle\Phi|H^{2n-i}|\Phi\rangle$.
Remaining roots of the polynomial correspond to the upper bounds of the excited-state energies. 
While both CMX and PDS methods are invariant under orbital rotations, the PDS approach, in contrast the CMX expansions, does not furnish size-extensive energies.

Recently, QC algorithms for  ITE
 and quantum Lanczos algorithm \cite{mcardle2019variational,motta2020determining,yeter2020practical} have attracted a lot of attention.  
One of the most appealing features of these methods is the avoidance of a large number of ancillae qubits and complex circuits. 
In this context, we believe that the CMX framework provides another way for the utilization of near-term quantum architectures, where one can almost entirely eliminate Trotter steps and significantly reduce the number of CNOT gates. Below we will describe a simple quantum algorithms for calculating moments $K_n=\langle\Phi|H^n|\Phi\rangle$ for an arbitrary trial state. \\
\section{Quantum algorithms for Connected Moments Expansion} 
In this Section we describe a general structure of the quantum algorithm for calculating moments $K_n$. Further simplifications and optimization of the circuit can be achieved by adding more ancillae qubits and/or merging with other quantum algorithms.

The main difference  between our algorithm  and ITE algorithms (for example QITE approach of Ref.\citenum{motta2020determining}) is the fact that in the latter approach the infinitesimal  ITE generated by unitary $H_j$, i.e. $e^{-\Delta_{\tau} H_j}$, is mirrored by a unitary evolution generated by the $e^{-i\Delta_{\tau}A_j}$ operator acting onto properly normalized states, while in the former method the standard moments $K_n=\langle\Phi|H^n|\Phi\rangle$ are directly calculated. 
Several quantum algorithms have been proposed for computing $K_n$. For example, since $H_n = i^n \frac{\partial^n }{\partial t^n} e^{\text{i}Ht} |_{t=0}$, Seki and Yunoki have shown \cite{seki2020quantum} that $K_n$ can be generally approximated by a linear combination of the time-evolution operators at $n+1$ different time variables. Also, to estimate ground state properties of a programmable quantum device, Kyriienko \cite{kyriienko2020quantum, bespalova2020hamiltonian} proposed a quantum inversion iteration algorithm where the inverse of Hamiltonian powers is decomposed as a sum of unitary operators. Here in this work, different from the above-mentioned approaches, we directly employ the Hadamard test to evaluate the contributions to $K_n$ for constructing connected moments and computing CMX energies.

\begin{figure}[ht]
\centering
\includegraphics[scale=0.23]{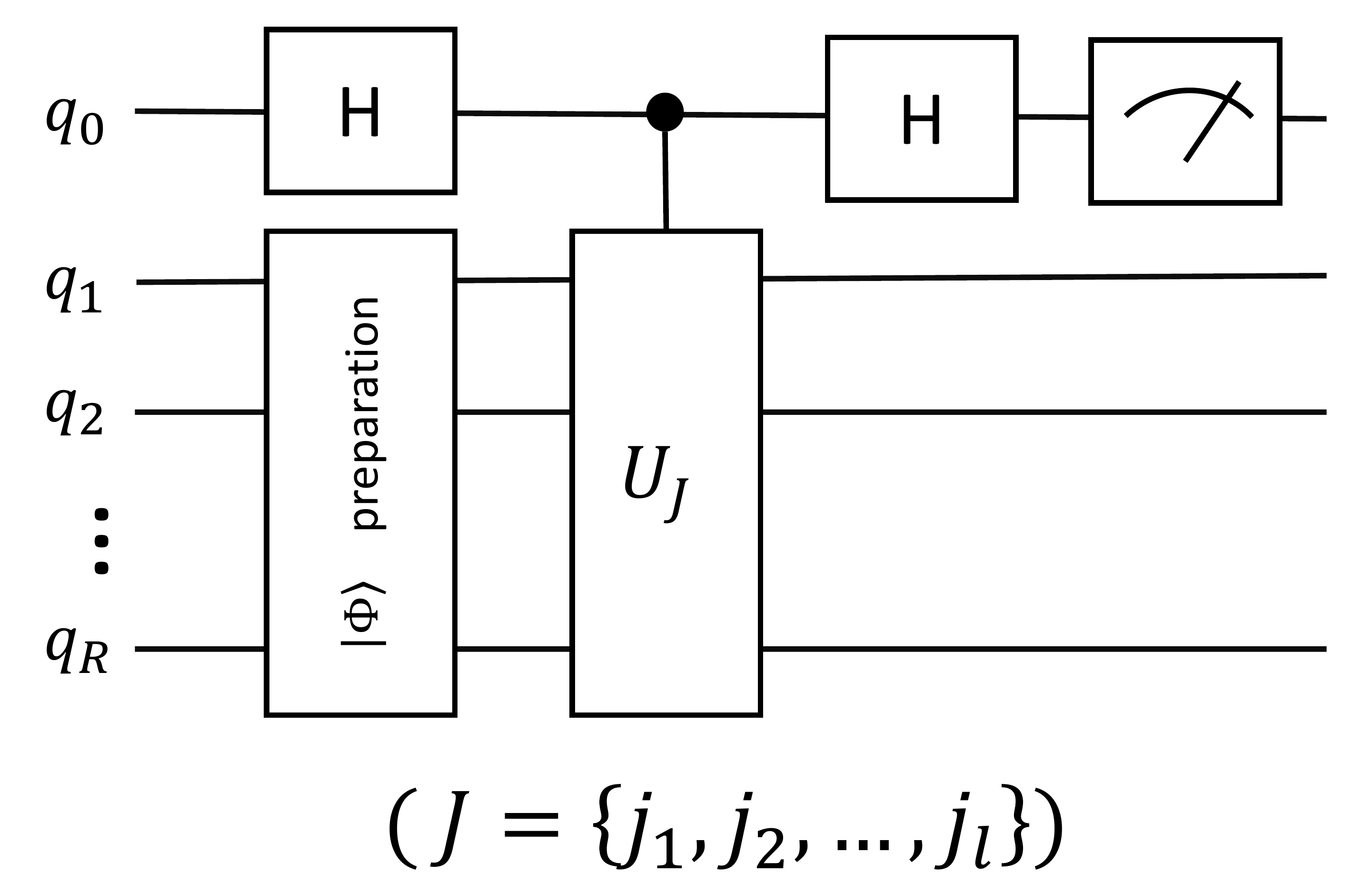}
\setlength{\abovecaptionskip}{-01pt}
\setlength{\belowcaptionskip}{-5pt}
\caption{
A schematic representation of the Hadamard test employed for calculating real part of
$\langle\Phi|U_J|\Phi\rangle$. 
The index $J$ corresponds to a string of indices $\lbrace j_1, j_2, \ldots, j_l \rbrace$). 
}
\label{fig1}
\end{figure}

\begin{figure}[ht]
\centering
\includegraphics[scale=0.29]{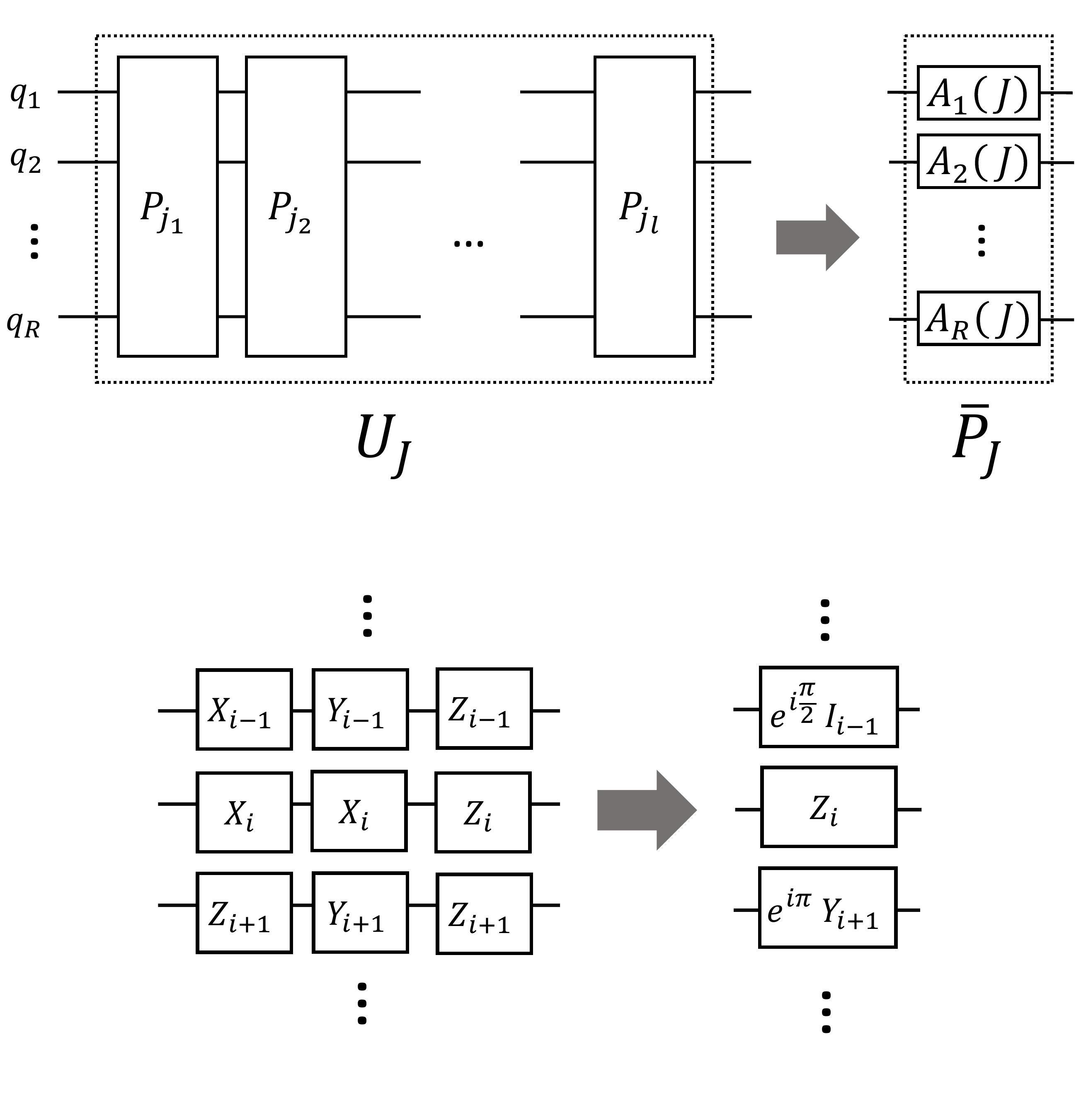}
\setlength{\abovecaptionskip}{-5pt}
\setlength{\belowcaptionskip}{-3pt}
\caption{
A schematic for the 
(Top panel) circuit-depth-reduction 
of the product of unitaries into a single unitary $\bar{P}_J$.
Single qubit gates 
$A_i(J)=e^{i\alpha_i(J)} u_i(J)$, where $e^{i\alpha_i(J)}$ is a phase factor and $u_i(J)$ corresponds either to the  identity operator ($I$) or one of three gates, $X_i$, $Y_i$, or $Z_i$
(for the sake of generality, we assume the presence of imaginary antisymmetric $Y$-type components in the Hamiltonian operator), 
(Bottom panel) a typical example of circuit-depth-reduction for $\langle\Phi|H^3|\Phi\rangle$.
}
\label{fig2}
\end{figure}

Quantum CMX (QCMX) algorithm works as follows: 
(1) preparing an initial state $|\Phi\rangle$.
In our numerical tests we used a simple single Slater determinant for $|\Phi\rangle$. 
Alternatively, one can envision the use of  ``non-aggressive" variant of VQE where a small number of  amplitudes that define basic static correlation effects in the sought-for-wave-function are optimized. 
(2) Performing Hadamard test to evaluate $p_J=\langle\Phi|U_J|\Phi\rangle$, where $U_J=\prod_{k=1}^{l} P_{j_k}$ (for CMX(K) formulation $l< 2K-1$) (see Fig. \ref{fig1}) and cumulative index $J$ designates l-tuple
$\lbrace j_1, \ldots, j_l \rbrace$.
The contribution from $p_J$ to $K_l$ is equal to 
$p_J \times \prod_{k=1}^{l}h_{j_k}$. 
(3) Compute connected moments $I$'s from moments $K$'s. 
(4) Once all  $I_l$ ($l=1,\ldots,2K-1$) are known we choose the optimal form of the CMX expansion. For example, if $I_3$ is close to 0, then we choose a CMX form that is not using $I_3$ in the denominator (this will provide an optimal utilization of the information obtained from QC). 
Additionally,  using identity $\sigma_i \sigma_j = \delta_{ij} I + i \epsilon_{ijk} \sigma_k\; (i,j,k=x,y,z)$ one can reduce entire $U_J$
to an effective unitary $\bar{P}_J$ corresponding to a tensor product of Pauli matrices and/or identity matrices with an appropriate product of phase factors (Fig. \ref{fig2}).
It means that the gate depth is exactly the same irrespective of the rank $l$ of the calculated moment  $K_l$ (of course, the number of terms to be evaluated in this way increases with the rank increasing).
Summarizing, the gate depth of the quantum CMX algorithm is mainly determined by the wave function preparation step and the practical realization of multi-qubit CNOT in the Hadamard test. In the case where the state preparation corresponds to flipping states from $|0\rangle$
to $|1\rangle$ using $X_i$ operators 
(for quantum registers corresponding to occupied spin-orbitals),
the resulting gate depth is extremely shallow. The utilization of VQE for $|\Phi\rangle$ state preparation increases the gate depth.
Even though, if it includes only a small number of amplitudes, the resulting gate depth should not be excessive. 
Moreover, employing trial wave function encapsulating basic correlation effects may have a positive effects on the accuracy of finite-order CMX formulations. When entangled trial wave function is used one can also see  advantages of using quantum computers - complexity and numerical overhead  of the expressions defining  expectation values of $H^n$ operators for correlated trial wave function(s) grow rapidly  
making classical calculation infeasible.

A large number of measurements is a common bottleneck of quantum algorithms, including VQE, ITE, and QCMX formalisms. Due to its simplicity, for the QCMX formalism, a simple reduced-measurement variant can be proposed. 
For example, one can select only the leading terms in the Hamiltonians' second-quantized form and perform the Fermion-to-qubit mapping, which results in the Hamiltonian $H’$ characterized by much smaller number of terms, i.e., 
\begin{equation}
H’=\sum_{j=1}^{M_R} h(R)_{j} P(R)_{j}
\end{equation}
where $M_R < M$. One should also notice that a single measurement of  specific matrix element
\begin{equation}
\langle\Phi(\theta)|P(R)_j|\Phi(\theta)\rangle \;,
\label{ubu1}
\end{equation}
(where for generality, we assume that the trial wave function can be optimized with respect to some variational parameters $\theta$)
can be used to determine  classes of contributions to moments of arbitrary rank. For example, for “diagonal” products, one has:
\begin{widetext}
\begin{eqnarray}
\hspace*{-0.5cm}  h(R)_{j}  \langle\Phi(\theta)|P(R)_j|\Phi(\theta)\rangle  & \longrightarrow &
\langle\Phi(\theta)|H|\Phi(\theta)\rangle  \;,\label{m1} \\
\hspace*{-0.5cm}  h(R)_{j}^2 \langle\Phi(\theta)| P(R)_j P(R)_j |\Phi(\theta)\rangle =  h(R)_{j}^2
&\longrightarrow & \langle\Phi(\theta)|H^2|\Phi(\theta)\rangle  \;,\label{m2} \\
\hspace*{-0.5cm}  h(R)_{j}^3 \langle\Phi(\theta)| P(R)_j P(R)_j  P(R)_j|\Phi(\theta) \rangle =  h(R)_{j}^3
\langle\Phi(\theta)|P(R)_j|\Phi(\theta)\rangle
&\longrightarrow & \langle\Phi(\theta)|H^3|\Phi(\theta)\rangle  \;,\label{m3}  \\
\hspace*{-0.5cm}  \ldots&&  \;,\nonumber 
\end{eqnarray}
\end{widetext}
where we used the fact that 
$\langle\Phi(\theta)| P(R)_j P(R)_j |\Phi(\theta)\rangle=1$
for normalized $|\Phi(\theta)\rangle$.
In a similar way one can calculate contributions from $\langle\Phi(\theta)|P(R)_i P(R)_j|\Phi(\theta)\rangle$, 
\linebreak
i.e. $\langle\Phi(\theta)|P(R)_i P(R)_j P(R)_i P(R)_j|\Phi(\theta)\rangle$, 
 $\langle\Phi(\theta)|P(R)_i P(R)_j P(R)_i P(R)_jP(R)_i P(R)_j|\Phi(\theta)\rangle$, 
etc. 
When combined, these two techniques can significantly reduce the number of measurements and enable a simple evaluation of higher-order moments
(see Supplementary Materials for more discussion). In the  quantum CMX extensions, in order  to reduce the effort associated with calculating $\langle\Phi|H^n|\Phi\rangle$ we are also  planning to utilize quantum power methods introduced in Ref.\citenum{seki2020quantum}.

\section{Simulations and results}
\begin{figure}[ht]
\centering
\includegraphics[scale=0.32]{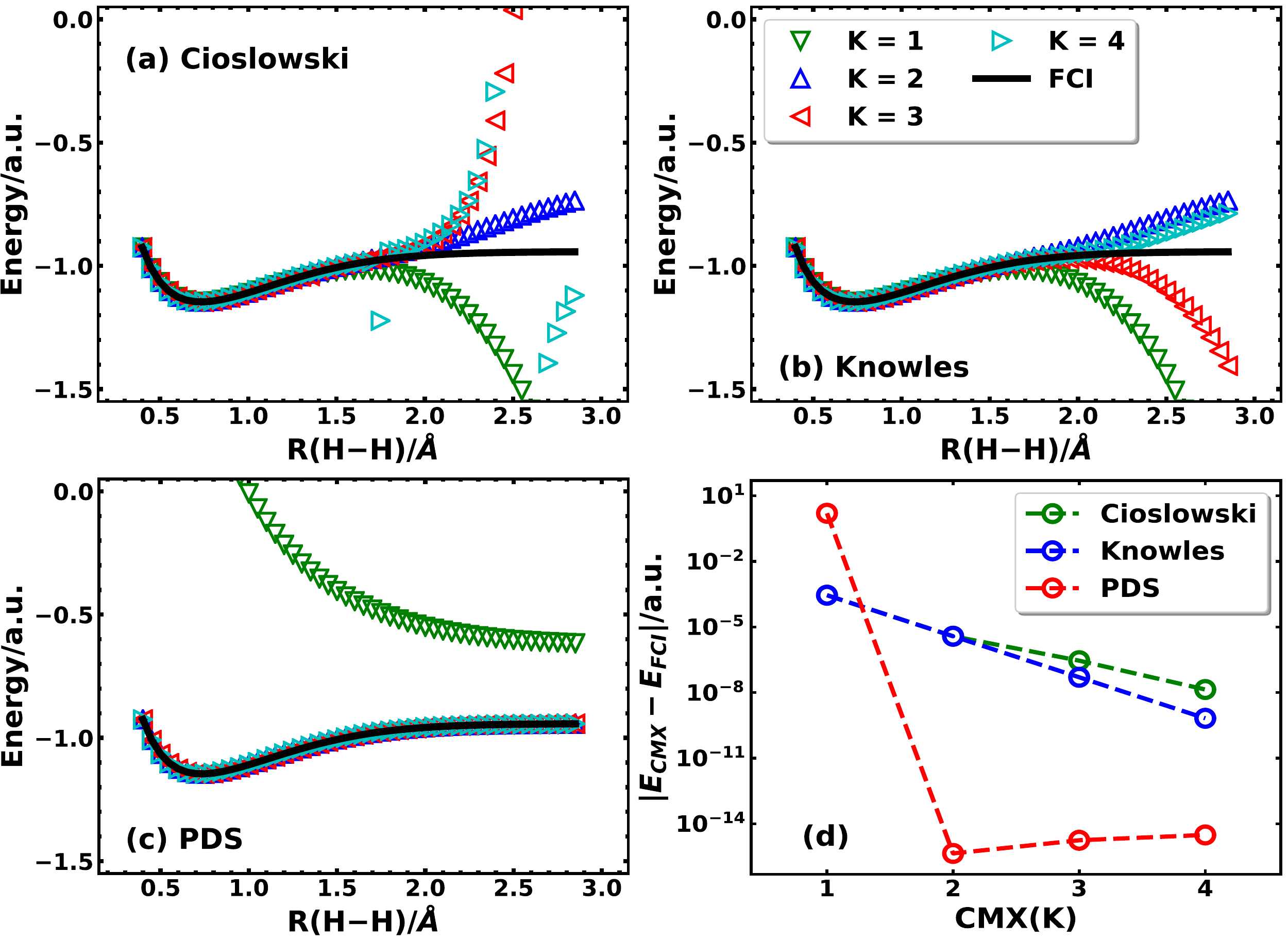}
\caption{
CMX(K) (K$=1-4$) results for
the H$_2$ molecule potential energy surface in minimum basis using three CMX variants, (a) Cioslowski, (b) Knowles, and (c) PDS. Their convergence behaviors as functions of expansion order (K) are collected in (d), where $R_{\text{H}-\text{H}}$ is fixed as 0.75 $\AA$, and the corresponding FCI energy (including nuclei repulsion) is $-1.145629458823643$ a.u. In all of the H$_2$ calculations, the H$_2$ system is represented by only two qubits with a trial vector $| 01 \rangle$ being universally applied, and the electronic Hamiltonian is effectively represented by a six-term Bravyi-Kitaev expression, $H^{\text{BK}} = g_0 I + g_1 Z_0 + g_2 Z_1 + g_3 Z_0 \otimes Z_1 + g_4 X_0 \otimes X_1 + g_5 Y_0 \otimes Y_1$ where the $R_{\text{H}-\text{H}}-$dependent real scalars, $g_i$ ($i=0-5$), are from Ref. \citenum{omalley2016scalable}. 
The fidelity of the trial vector with respect to the true ground state, and the second-quantized form of the electronic Hamiltonian (and its relation to two-site Hubbard model) are given Supplementary Materials}.
\label{fig3}
\end{figure}

\begin{figure}[ht]
\centering
\includegraphics[scale=0.32]{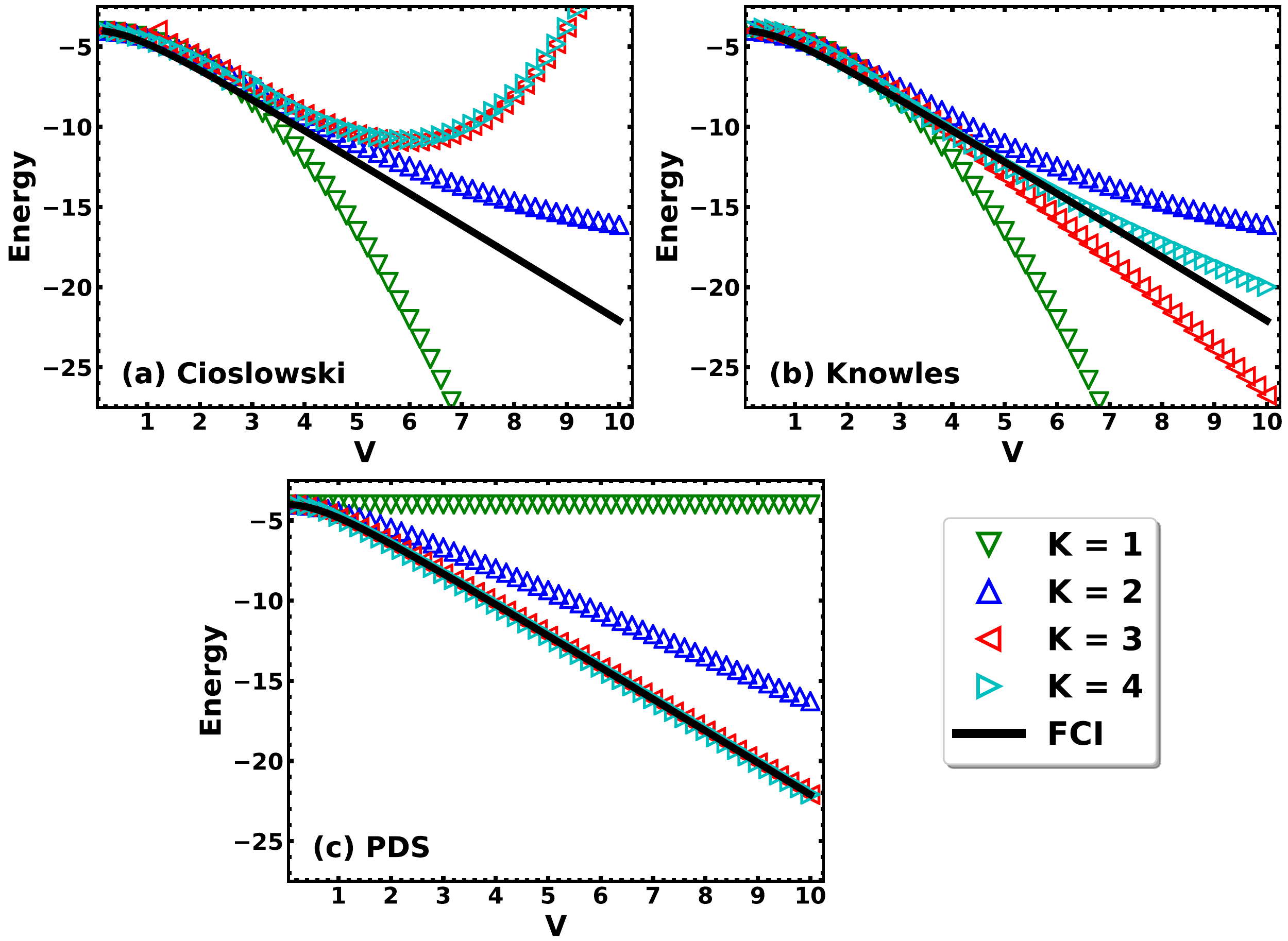}
\caption{
CMX(K) (K$=1-4$) results for the total energy of the single impurity Anderson model (SIAM) as a function of hybridization strength (V) using three CMX variants, (a) Cioslowski, (b) Knowles, and (c) PDS. In all of the SIAM calculations, the system is represented by four qubits with a trial vector $|0110\rangle$ being universally applied, and the Hamiltonian for the two-site (i.e. one impurity site and one bath site) half-filling Anderson impurity model is represented by its Jordan-Wigner expression, $H^{\text{JW}}_{\text{SIAM}} = \frac U 4 (I - Z_1) \otimes (I - Z_3) + \frac {\epsilon_0 - \mu}{2} (2I - Z_1 - Z_3) + \frac{\epsilon_1 - \mu}{2}( 2I - Z_2 - Z_4) + \frac V 2 (X_1\otimes X_2 + Y_1\otimes Y_2 + X_3\otimes X_4 + Y_3\otimes Y_4)$, where for half-filling  we fix the chemical potential $\mu$ as half of the local Hubbard repulsion $U$ , and the impurity and bath site energies respectively as $\epsilon_0 = 0$ and $\epsilon_1 = \mu$. $V$ is the hybridization that allows the hopping between bath and impurity sites. The analytical FCI energy expression under this condition is $-\frac 1 4 (U + \sqrt{U^2 + 64 V^2})$.
}
\label{fig4}
\end{figure}

For building  quantum circuits to calculate the moments, $K_l$'s, we used Qiskit software.\cite{Qiskit} As benchmark systems we chose H$_2$ molecule in minimum basis (for various geometries corresponding  to situations characterized by weak and strong correlation effects) \cite{omalley2016scalable}
 and two-site single-impurity Anderson model (for a broad range of hybridization strength $V$) described by model Hamiltonians. 
 In our studies we used various orders of Cioslowski CMX(K), Knowles CMX(K),\cite{knowles1987validity} and K-th order Peeters, Devreese, Soldatov expansion, PDS(K). 
 In our quantum simulations, individual spin orbitals are directly assigned to qubits. For each qubit, $|1\rangle$ and $|0\rangle$ correspond to occupied and unoccupied states of the spin orbital. We applied simple $X$ gates on vacuum state to generate our trial states. Fidelities of trial states are given in the Supplementary Materials.
 The results of quantum simulations are shown in Figs. \ref{fig3} and \ref{fig4}.
%

In Fig. \ref{fig3}, all three CMX variants with up to fourth order expansions are able to reproduce the FCI energy in at least mHartree level for $R_{\text{H}-\text{H}}$ being up to 1.50 $\AA$. The only exception is PDS(1), which corresponds to  $I_1 = \langle \Phi | H | \Phi \rangle$.  The higher-order PDS expansions perform exceptionally well, and the deviations between the PDS(K) (K$=2-4$) and FCI energy are below 10$^{-14}$ a.u. The original Cioslowski CMX results start to diverge and show singularity on the potential energy surface when $R_{\text{H}-\text{H}}$ is approaching dissociation limit ($R_{\text{H}-\text{H}} > 1.50 \AA$). 

The singularity problem originates from the near-zero connected moments defining the denominators in the  CMX energy expansion (see Eqns. (\ref{cmx2}) and (\ref{cmx3})).
Mancini {\it et al.}\cite{mancini1994analytic} have shown how to mitigate these singularities by re-summing the CMX expansion (i.e. the AMX approach) and introducing new class of denominators corresponding to non-zero higher order moments. 
By a similar substitution derived from a Pad\'{e} approximant, Knowles \cite{knowles1987validity} has 
shown that the singularity could be largely eliminated. This can be observed from H$_2$ 
results as shown in Fig. \ref{fig3}a,b, where at the dissociation limit, the Knowles's approach (Fig. \ref{fig3}b) 
is able to provide a smooth curve and move towards the FCI limit in an oscillatory manner, while divergences appears in the original Cioslowski's CMX formalism after the second order, and additionally the singularity emerges in the original CMX(4) for $R_{\text{H}-\text{H}}$ being $\sim$1.75 $\AA$ and $\sim$2.50 $\AA$.
Applications of QCMX to  H4 system are included in the Supplementary Material.

We further examine the performance of the CMX variants in terms of computing the ground state energy ranging from weak to strong correlation in the context of a single-impurity Anderson model (SIAM). As can be seen from Fig. \ref{fig4}, for fixed Hubbard repulsion $U$ ($U=8$), the exact correlation energy curve monotonically declines as the hybridization $V$ becomes large. Among the low order CMX(K) (K$=1,2$) results, all the curves behave very similarly showing the same trend as the FCI curve (except the PDS(1) curve where the correlation contribution from the single-particle $V$ terms operating on the present trial wave function is zero). Among high order CMX(K) (K$>2$) results, the original CMX diverges as the $V$ becomes large, while the Knowles's approach improves the original CMX results by showing slow convergence as the expansion order increases. On the other hand, PDS approach shows fast convergence, and basically reproduces the FCI results after the second order. 

\begin{figure}[ht]
\centering
\includegraphics[scale=0.32]{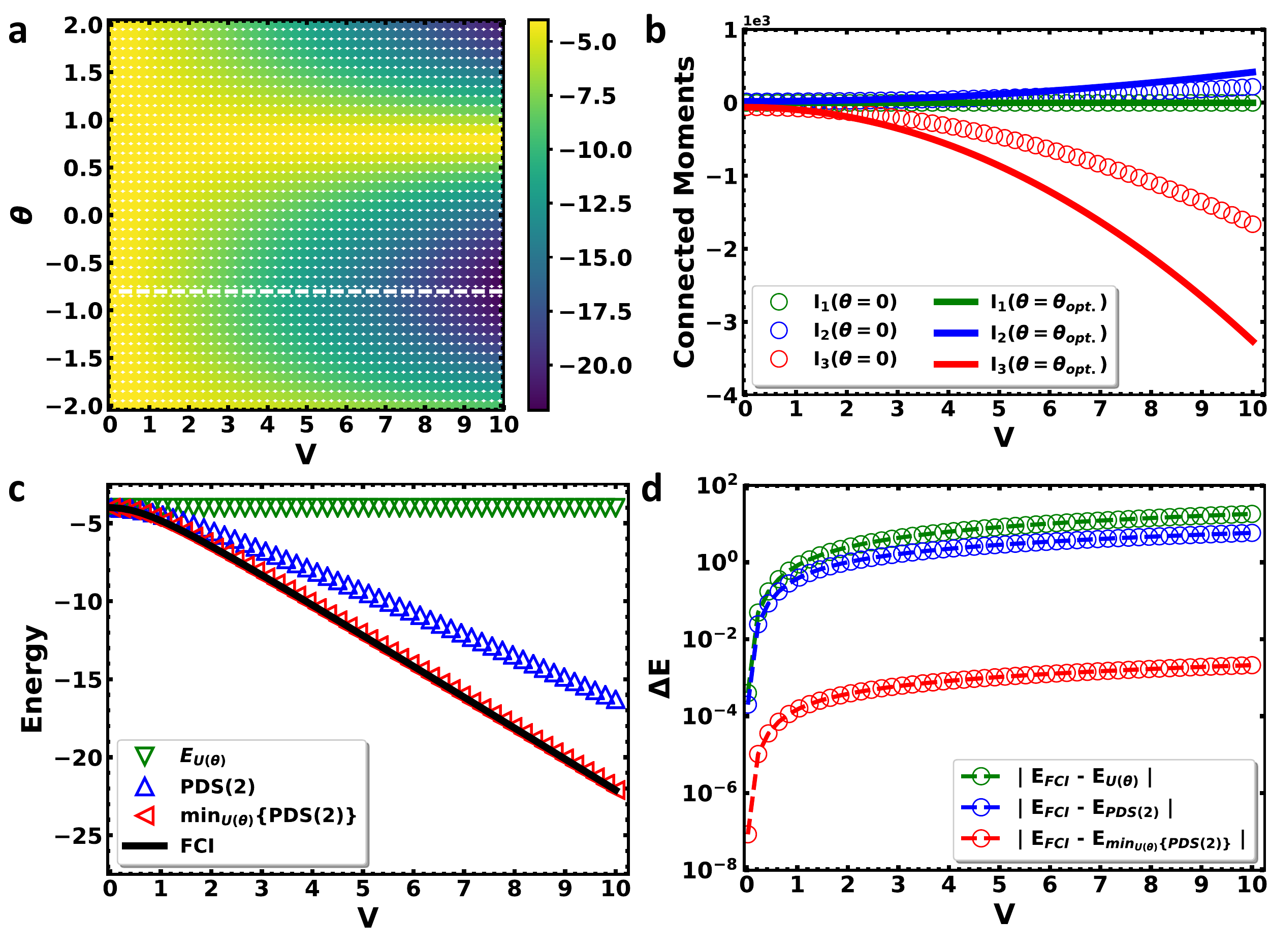}
\caption{(a) PDS(2) energy as a function of unitary rotation angle ($\theta$) and hybridization strength ($V$), where the white dashed line traces the minimum PDS(2) energy for a given $V$ (here $\theta_{\text{opt.}}\sim-0.8$ for all the $V$ values). (b) The changes of the connected moments up to the third order as functions of $V$ value at $\theta=0$ and $\theta = \theta_{\text{opt.}}$ respectively. (c) Ground state energy curves computed from unitary rotation $U(\theta)$ (i.e.   $E_{U(\theta)} = \langle \Phi | U(\theta)^\dagger H^{\text{JW}}_{\text{SIAM}} U(\theta) | \Phi \rangle$), PDS(2), and minimizing PDS(2) through rotating the trial wave function using $U(\theta)$ ($\min_{U(\theta)}\{\text{PDS}(2)\}$). (d) Deviations of the computed ground state energies w.r.t. the full CI energies employing the three approaches mentioned in (c) over the studied $V$ range. As suggested in Ref. \citenum{mccaskey2019quantum}, the unitary rotation is defined as $U(\theta) = \text{exp}(\text{i}\theta Y_0X_1X_2X_3)$. The trial wave function is $|0110\rangle$. 
The fidelities of the trial wave function before and after the unitary rotation with respect to the true ground state are given Supplementary Materials.
}
\label{fig5}
\end{figure}

Note that in all the CMX calculations shown in Figs. \ref{fig3} and \ref{fig4}, a simple form of the trial wave function (i.e. single determinant wave function) is used in either weak or strong correlation scenarios, and the commonly used unitary coupled-cluster ansatze, or in general the unitary operations for the state preparation as discussed in the previous practices (see e.g. Refs. \citenum{omalley2016scalable,kandala2017hardware,hempel2018quantum,mccaskey2019quantum,Keen2020quantum}), were not invoked. In other word, a crude trial wave function may be still useful by employing the proposed quantum algorithm to obtain highly accurate energies. This does not defy the importance of the state preparation at the dissociation or strong correlation limits, but rather provides an alternatively great simplification for consideration when dealing with some quantum applications. Remarkably, different from the other two CMX variants, the PDS is in principle able to provide upper bounds for all the energy levels,\cite{soldatov1995generalized} and the upper bound for the ground state energy from the PDS approach is relatively tight even at the dissociation and strong correlation limits. Therefore, the PDS can further work with some unitary parameterization to improve the accuracy of the ground state energy computed at the low expansion order. As shown in Fig. \ref{fig5}, a unitary parameterization suggested in Ref. \citenum{mccaskey2019quantum}, $U(\theta) = \text{exp}(\text{i}\theta Y_0X_1X_2X_3)$, can 
improve the fidelity of the trial wave function (see Supplementary Materials), thus
help the PDS(2) to well reproduce the FCI curve. As can be seen in Fig. \ref{fig5}d, the energy deviation can be reduced on average by about three orders of magnitude when 
variational 
PDS(2) energies are minimized through rotating the trial wave function using the selected unitary operation $U(\theta)$ over the studied range of $V$ values. Note here that employing the selected unitary parametrization alone cannot produce FCI energy for the four-qubit, 2-electron problem (here $E_{U(\theta)} = I_1 = \langle \Phi | U(\theta)^\dagger H^{\text{JW}}_{\text{SIAM}} U(\theta) | \Phi \rangle = -4$ for $|\Phi\rangle = |0110\rangle$ and arbitrary $\theta$, see Fig. \ref{fig5}(b,c)). Similar results can be obtained with alternative parametrizations of $U(\theta)$ given by $\text{exp}(\text{i}\theta X_0Y_1X_2X_3)$ or $\text{exp}(\text{i}\theta Y_0X_1)$.
We have performed the CMX/PDS simulation for a larger systems. For example, in Supplementary Materials, we have shown the CMX/PDS quantum simulations using up to 12 qubits are able to target the full CI energies of the singlet and triplet states of H$_4$ system described by different geometries and active spaces of various sizes, especially for the strongly correlated planar H$_4$ system.

We have also preliminarily studied the noise impact on the ideal CMX results employing the noise models that adopt reported values of Rigetti Aspen-1 QPUs and typical parameters for benchmarking error mitigation algorithms.\cite{temme17_180509,motta2020determining} The results are shown in Fig. \ref{fig6} and \ref{fig7}. As can be seen, the noise models slightly shifted the CMX(2)/PDS(2) results with respect to the idealized ones. Remarkably, for H$_2$ at larger $R_{H-H}$ noise can ``regularize" idealized CMX(2) curve when the latter starts to disclose singularity due to the close-to-zero $I_3$ moments.
For Anderson model, even though the effect of noise is slightly bigger in the simulation with the correlated trial wave function, the quality of noisy PDS(2) results with correlated reference is still better than the noisy PDS(2) results employing  $|0110\rangle$ reference. Note that we do not include all sources of error in our emulation, and more comprehensive testing will be pursued in our future work.

The striking advantage offered by quantum CMX algorithms is its ``insensivity" with respect to  the choice of initial state. As long as the initial state is non-orthogonal to the exact state, the CMX expansions (especially the PDS approach) are capable of reproducing exact or near-to-exact energies. This feature could largely eliminate the challenging state-preparation process in quantum algorithms. As we can see from the performance of the CMX variants in Figs. \ref{fig3} and \ref{fig4}, instead of constructing a better trial wave function, the 
moments matrix  ${\bf M}$   can be  constructed to avoid singularity problems for quasi-degenerate  situations. 
Finally, it is worth mentioning that based on the same Krylov subspace we have also performed the Lanczos approach (for up to 4-th order) for the present studied systems, and did find any observable difference between PDS(K) and Lanczos(K) results. However, noticing the numerical performance differences between different eigensolvers (or linear solvers) that are based on a same Krylov subspace in the classical computing, as well as previous classical comparisons between CMX and Lanczos results,\cite{knowles1987validity, prie1994relation, mancini1994analytic, ullah1995removal, mancini1995avoidance, fessatidis2006generalized, fessatidis2010analytic} we suppose performance differences between PDS and Lanczos methods would manifest for certain systems at certain scenarios. We believe that PDS and Lanczos methods might be complementary to each other for challenging many-body QC studies. Extensive testing for other models and molecular systems are currently underway.

{\color{blue}
%
\begin{figure*}[ht]
\centering
\includegraphics[width=\textwidth]{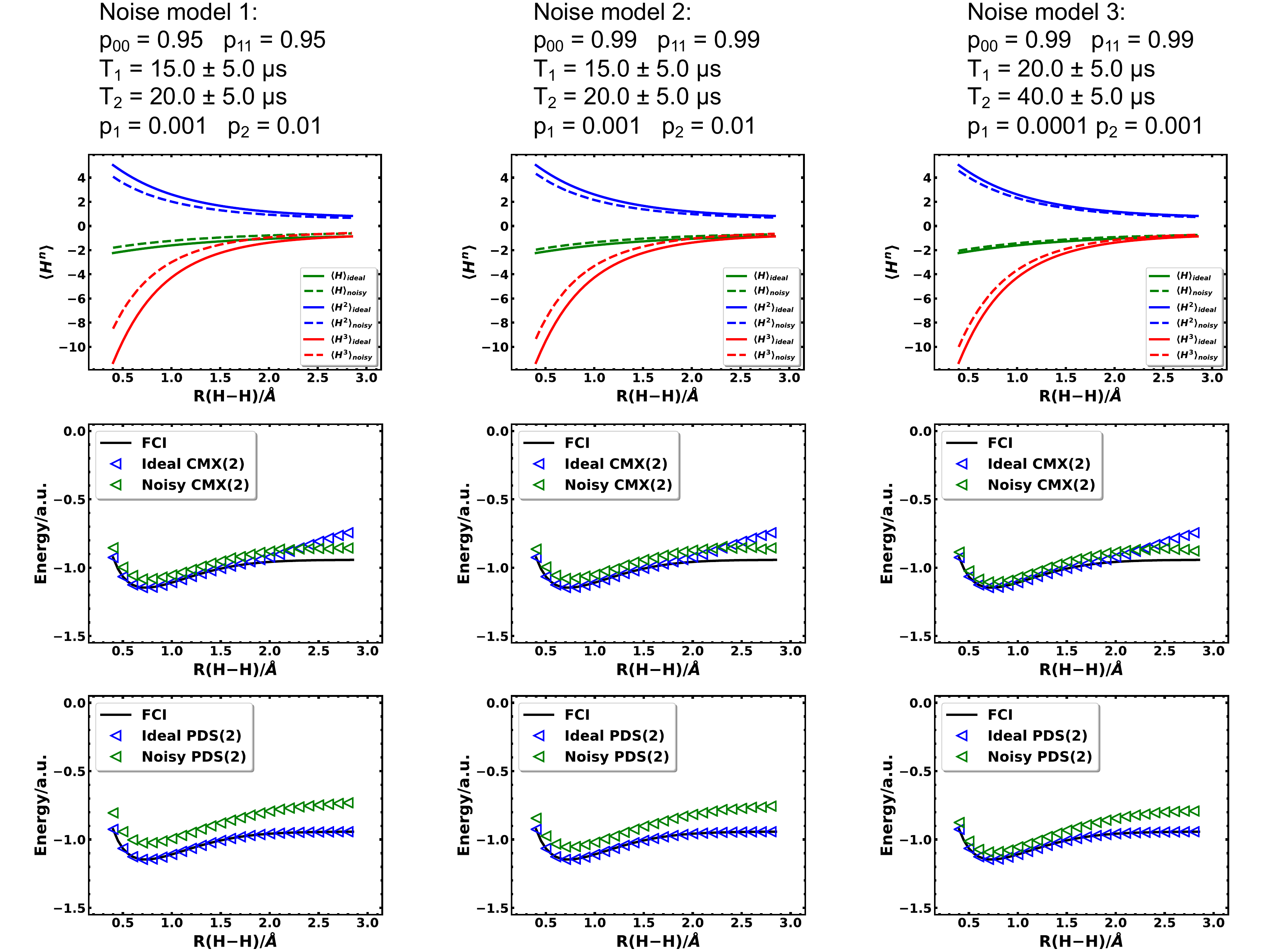}
\caption{Noisy CMX(2) and PDS(2) simulations for potential energy surface (PES) of H$_2$ molecule.
Three noise models were employed to study the impacts of 1-qubit readout error, 1/2-qubit thermal relaxation error, and 1/2-qubit depolarizing error
on the CMX(2) and PDS(2) simulations of H$_2$ PES. The 1-qubit readout error is characterized by the probability of recording a measurement
outcome of 0 given the state is in $|0\rangle$ ($p_{00}$) and the probability of recording a measurement
outcome of 1 given the state is in $|1\rangle$ ($p_{11}$). The thermal relaxation error accounts for the qubit environment, and the error rates on instruction
are parameterized by a thermal relaxation time constant $T_1$, a dephasing time constant $T_2$, and gate times. In the present test, the gate times
for $U_1$, $U_2$, $U_3$, and $Controlled$-$X$ gates were fixed to 0 $ns$, 50 $ns$, 100 $ns$, and 300 $ns$, respectively. We also fixed the reset time and measure time to be 1 $\mu s$.
The $T_1$ and $T_2$ are sampled from normal distributions centered at the designated times with a standard deviation of 5 $\mu s$.
The depolarizing error is characterized by the depolarizing probability for 1-qubit gates ($p_1$) and 2-qubit gates ($p_2$).
The expectation values of Hamiltonian powers (up to $\langle H^3\rangle$) with and without noise models are also given. The CMX(2) and PDS(2) energy values (with and without
noise models) are collected in Tabs. S1 and S2 in Supplementary materials.}
\label{fig6}
\end{figure*}
\begin{figure}[ht]
\centering
\includegraphics[scale=0.25]{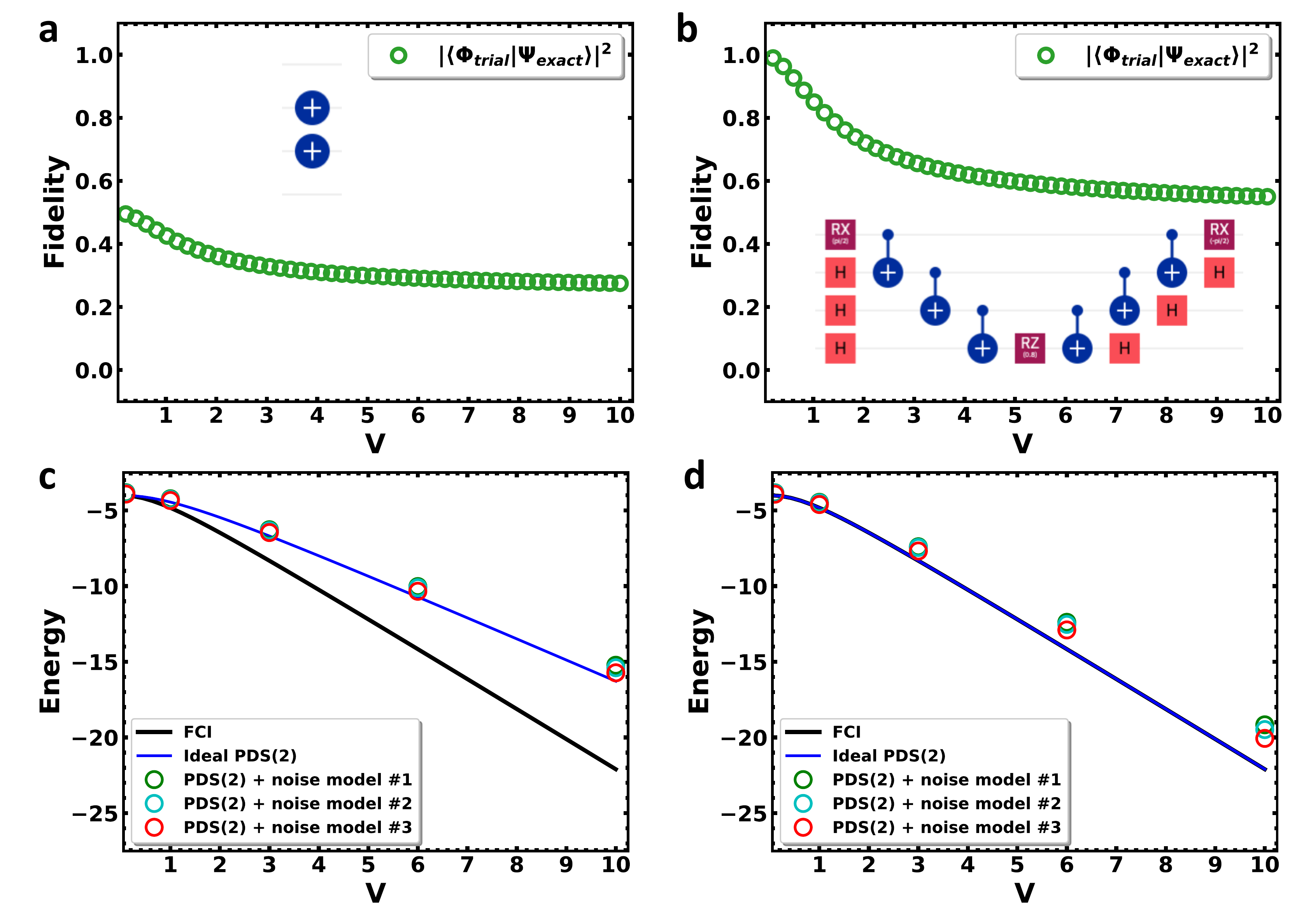}
\caption{ Noisy PDS(2) simulations for the Anderson model employin $|0110\rangle$ as a trial wave function (left panel) and $\text{exp}(\text{i}\theta X_0Y_1X_2X_3)|0110\rangle$ with optimal value of $\theta $ parameter (right panel). (a,b) Fidelity of trial states employed in the PDS(2) simulations. The circuits used to generated the trial states are given in the insets. The circuit in (a) corresponds to bit-flips at the second and third qubits that generates the state $|0110\rangle$. The circuit in (b) corresponds to a unitary rotation, $U(\theta)=\exp(\text{i} \theta Y_0X_1X_2X_3)$ ($\theta=0.8$), on $|0110\rangle$ state, in which the rotations in the two $RX$ gates are $\frac {\pi}{2}$ and $-\frac {\pi}{2}$ respectively. The optimized rotation in the $RZ$ gate is 0.8. The noise models are same as those used for H$_2$ as shown in Fig. \ref{fig6}. The PDS(2) simulations with noise models were performed at $V=0.1, 1, 3 ,6, \text{and } 10$. For the half-filling impurity calculations, we fix the local Hubbard repulsion $U=8$, the chemical potential $\mu=4$, the impurity site energy $\epsilon_0 =0$, and the bath site energy $\epsilon_1=4$. Numerical PDS(2) energy values are collected in Tab. S3. Fidelities of the trial states used in the simulation are given with respect to exact ground state.}
\label{fig7}
\end{figure}
} 

\section{Conclusions}
We demonstrated the feasibility of a new QC algorithm based on calculating various types of CMX expansions. The discussed algorithm is 
robust in the sense of possible gate depth reduction 
and could be highly scalable with the increasing number of available qubits. Since the CMX algorithms offer a trade-off between the quality of the trial wave function and the rank of the moments required to achieve a high level of accuracy, we believe that its combination with the VQE approach, that is utilizing the unitary coupled-cluster representation of the trail wave function,
may provide a much needed formulation for improving the quality of  VQE energies. Additionally, the flexibility of CMX re-summation techniques allows one to define customized (or optimal)  form of the expansion that avoids possible  singularities associated with the near-zero values of the calculated connected moments. In our studies, we demonstrated that the PDS expansions
provide superior results compared to other CMX expansions for the systems considered here, especially for the strongly correlated situations (characterized by large $R_{\text{H}-\text{H}}$ distance or large hybridization strength) studied with a relatively poor choice of trial wave function corresponding to a single Slater determinant.  
Our studies also  indicate that  the flexibility in choosing of the reference function may be a regulatory factor to establish a compromise between accuracy and the effect of noise. 
In our future studies we will explore the applicability of the PDS expansions for describing multi-configuration states and excited states. Of special interest will be integrating quantum CMX algorithms 
with downfolded Hamiltonians \cite{bauman2019downfolding,downfolding2020t} and  tensor decomposition techniques for defining effective interactions\cite{peng2017highly,motta2018low,motta2019efficient,von2020quantum}.
We will also explore the effect of  noise in  CMX/PDS simulations for larger systems.

\section{Supplementary Material}
See supplementary material for the  second quantized form of  the H$_2$ and SIAM Hamiltonians, , results of noisy CMX executions for H$_2$ and SIAM models, fidelity analysis of the trial state, and results for H$_4$ systems obtained  in simulations with and without  the noise.

\section{Acknowledgement}
This work was supported by  the ``Embedding QC into Many-body Frameworks for Strongly Correlated  Molecular and Materials Systems'' project, which is funded by the U.S. Department of Energy(DOE), Office of Science, Office of Basic Energy Sciences, the Division of Chemical Sciences, Geosciences, and Biosciences.
All claculations have been perfromed at the Pacific Northwest National Laboratory (PNNL).
PNNL is operated for the U.S. Department of Energy by the Battelle Memorial Institute under Contract DE-AC06-76RLO-1830.

\section{Data Availability}
The data that support the findings of this study are available
from the corresponding author upon reasonable request.



%

\end{document}